# A Simulation Model for Evaluating Distributed Systems Dependability

Ciprian Dobre, Florin Pop, Valentin Cristea
*Faculty of Automatics and Computer Science, University Politehnica of Bucharest, Romania*
E-mails: {ciprian.dobre, florin.pop, valentin.cristea}@cs.pub.ro

**Abstract**

*In this paper we present a new simulation model designed to evaluate the dependability in distributed systems. This model extends the MONARC simulation model with new capabilities for capturing reliability, safety, availability, security, and maintainability requirements. The model has been implemented as an extension of the multithreaded, process oriented simulator MONARC, which allows the realistic simulation of a wide-range of distributed system technologies, with respect to their specific components and characteristics. The extended simulation model includes the necessary components to inject various failure events, and provides the mechanisms to evaluate different strategies for replication, redundancy procedures, and security enforcement mechanisms, as well. The results obtained in simulation experiments presented in this paper probe that the use of discrete-event simulators, such as MONARC, in the design and development of distributed systems is appealing due to their efficiency and scalability.*

## 1. Introduction

Modeling and simulation were seen for long time as viable solutions to develop new algorithms and technologies and to enable the enhancement of large-scale distributed systems, where analytical validations are prohibited by the scale of the encountered problems. The use of discrete-event simulators in the design and development of large scale distributed systems is appealing due to their efficiency and scalability.

Together with the extension of the application domains, new requirements have emerged for large scale distributed systems; among these requirements, reliability, safety, availability, security and maintainability, in other words dependability [1], are needed by more and more modern distributed applications, not only by the critical ones.

However, building dependable distributed systems is one of the most challenging research activities. The characteristics of distributed systems make dependability a difficult problem from several points of view. The geographical distribution of resources and users that implies frequent remote operations and data transfers lead to a decrease in the system's safety and reliability and make it more vulnerable from the security point of view. Another problem is the volatility of the resources, which are usually available only for limited periods of time. The system must ensure the correct and complete execution of the applications even in the situations when the resources are introduced and removed dynamically, or when they are damaged. The management of distributed systems is also complicated by the constraints that the applications and the owners of the resources impose; in many cases there are conflicts between these constraints – for example, an application needs a long execution time and performs database operations, while the owner of the machine on which the application could be run only makes it available in a restricted time interval and does not allow database operations.

In this paper we present a simulation model designed to evaluate the dependability in distributed systems. The proposed model extends the MONARC simulation model [16] with new capabilities for capturing reliability, safety, availability, security, and maintainability requirements. The model has been implemented as an extension of the multithreaded, process oriented simulator MONARC, which allows the realistic simulation of a wide-range of distributed system technologies, with respect to their specific components and characteristics. The extended simulation model includes the necessary components to inject various failure events, and provides the mechanisms to evaluate different strategies for replication, redundancy procedures, and security enforcement mechanisms, as well. The paper extends the results presented in [10], introducing the simulation model designed for dependability of distributed systems. The results obtained in simulation experiments presented in this paper probe that the use of discrete-event simulators, such as MONARC, in the design and development of distributed systems is appealing due to their efficiency and scalability.

The rest of this paper is structured as follows. Section 2 presents related work in the field of modeling distributed systems, with a special accent on the evaluation of dependability. Next we present the MONARC architecture. The next sections present the simulation model being proposed, together with its implementation within the MONARC simulator. In section 6 we present the obtained results. Finally, in section 7 we present some conclusions and future work.

## 2. Related work

*SimGrid* [2] is a simulation toolkit that provides core functionalities for the evaluation of scheduling algorithms in distributed applications in a heterogeneous, computational Grid environment. It aims at providing the right model and level of abstraction for studying Grid-based scheduling algorithms and generates correct and accurate simulation results. *GridSim* [3] is a grid simulation toolkit developed to investigate effective resource allocation techniques based on computational economy. *OptorSim* [4] is a Data Grid simulator project designed specifically for testing various optimization technologies to access data in Grid environments. OptorSim adopts a Grid structure based on a simplification of the architecture proposed by the EU DataGrid project. *ChicagoSim* [5] is a simulator designed to investigate scheduling strategies in conjunction with data location. It is designed to investigate scheduling strategies in conjunction with data location.

None of these projects present general solutions to modeling dependability technologies for large scale distributed systems. They tend to focus on providing evaluation methods for the traditional research in this domain, which up until recently targeted the development of functional infrastructures. However, lately, the importance of dependable distributed systems was widely recognized and this is demonstrated by the large number of research projects initiated in this domain. Our solution aims to provide the means to evaluate a wide-range of solutions for dependability in case of large scale distributed systems.

Security in particular has never been properly handled by any of these projects before. The only currently existing simulator that offers the possibility to evaluate security solutions designed for distributed systems is G3S (Grid Security Services Simulator) [6]. It was developed so as to support various authentication mechanisms including X.509 certificates or Kerberos tickets and includes mechanisms for disseminating security threats, for evaluating various access control policies, etc. The simulator is based on the simulation model found in GridSim. We too support all the mechanisms found in G3S and some others. In addition we offer the possibility of evaluating security in a more general context, considering the entire context of distributed systems, with its specific characteristics.

An issue here is related to the generic evaluation of dependable distributed systems. A fault occurring in such systems could lead to abnormal behavior of any of the system's components. For this reason we argue that a correct evaluation of dependability in distributed systems should provide a complete state of the entire distributed system. Because of the complexity of the Grid systems, involving many resources and many jobs being concurrently executed in heterogeneous environments, there are not many simulation tools to address the general problem of Grid computing. The simulation instruments tend to narrow the range of simulation scenarios to specific subjects, such as scheduling or data replication. The simulation model provided by MONARC is more generic that others, as demonstrated in [7]. It is able to describe various actual distributed system technologies, and provides the mechanisms to describe concurrent network traffic, to evaluate different strategies in data replication, and to analyze job scheduling procedures.

## 3. MONARC Architecture

MONARC is built based on a process oriented approach for discrete event simulation, which is well suited to describe concurrent running programs, network traffic as well as all the stochastic arrival patterns, specific for such type of simulations [8]. Threaded objects or "Active Objects" (having an execution thread, program counter, stack...) allow a natural way to map the specific behavior of distributed data processing into the simulation program. However, as demonstrated in [9], because of the considered optimizations, the threaded implementation of the simulator can be used to experiment with scenarios consisting of thousands of processing nodes executing a large number of concurrent jobs or with thousands of network transfers happening simultaneously.

In order to provide a realistic simulation, all the components of the system and their interactions were abstracted. The chosen model is equivalent to the simulated system in all the important aspects. A first set of components was created for describing the physical resources of the distributed system under simulation. The largest one is the regional center (Figure 1), which contains a site of processing nodes (CPU units), database servers and mass storage units, as well as one or more local and wide area networks. Another set of components model the behavior of the applications and their interaction with users. Such components are the "Users" or "Activity" objects which are used to generate data processing jobs based on different scenarios.

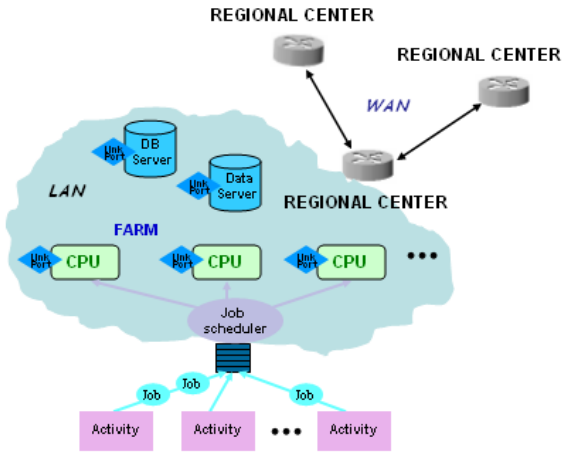

*Figure 1. The Regional center model.*

The job is another basic component, simulated with the aid of an active object, and scheduled for execution on a CPU unit by a "Job Scheduler" object. Any regional center can dynamically instantiate a set of users or activity objects, which are used to generate data processing jobs based on different simulation scenarios. Inside a regional center different job scheduling policies may be used to distribute jobs to corresponding processing nodes. One of the strengths of MONARC is that it can be easily extended, even by users, and this is made possible by its layered structure. The first two layers contain the core of the simulator (called the "simulation engine") and models for the basic components of a distributed system (CPU units, jobs, databases, networks, job schedulers etc.); these are the fixed parts on top of which some particular components (specific for the simulated systems) can be built. The particular components can be different types of jobs, job schedulers with specific scheduling algorithms or database servers that support data replication. The diagram in Figure 2 presents the MONARC layers and the way they interact with a monitoring system. In fact, one other advantage that MONARC have over other existing simulation instruments covering the same domain is that the modeling experiments can use real-world data collected by a monitoring instrument such as MonALISA, an aspect demonstrated in [11]. This is useful for example when designing experiments that are meant to experiment new conditions starting from existing real distributed infrastructures.

Using this structure it is possible to build a wide range of models, from the very centralized to the distributed system models, with an almost arbitrary level of complexity (multiple regional centers, each with different hardware configuration and possibly different sets of replicated data).

The maturity of the simulation model was demonstrated in previous work. For example, a number of data replications experiments were conducted in [8], presenting important results for the future LHC experiments, which will produce more than 1 PB of data per experiment and year, data that needs to be then processed. A series of scheduling simulation experiments were presented in [8], and [12].

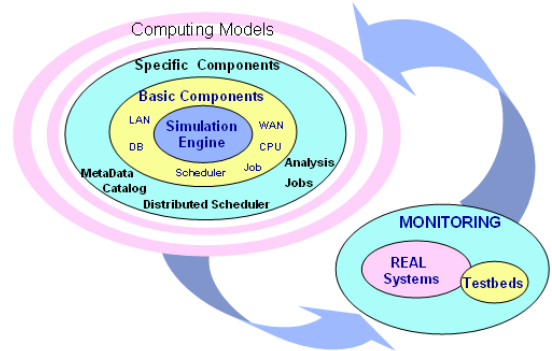

*Figure 2. The layers of MONARC.*

In [10] we presented an extension to the model designed to simulating fault tolerance in distributed systems using MONARC. The solution was able to model failures in distributed systems at hardware level (abnormalities in the functionality of hardware components) or software level (the middleware or application deviating from their normal functionality or delivery of services). In this we extended on this, implementing the proposed model in MONARC, evaluating it, but also adding additional mechanisms (security, different types of failures, at hardware and software levels, to model their occurrences and detection, as well as recovery and masking mechanisms) to cover a complete set of dependability characteristics, in its generic sense.

The characteristics of large scale distributed systems make the problem of assuring dependability a difficult issue because of several aspects. A first aspect is the geographical distribution of resources and users that implies frequent remote operations and data transfers; these lead to a decrease in the system's safety and reliability and make it more vulnerable from the security point of view. Another problem is the volatility of the resources, which are usually available only for limited periods of time; the system must ensure the correct and complete execution of the applications even in situations such as when the resources are introduced and removed dynamically, or when they are damaged.

In [10] we proposed an extension to the MONARC's model. In this paper we present the complete model designed to consider all aspects of dependability. Figure 3 presents the components of the dependable modeling layer. The extension to the simulation model relates to modeling faults appearing inside the modeled distributed system. In a distributed system failures can be produced at hardware level (abnormalities in the functionality of hardware components) or software level (the middleware or application deviating from their normal functionality or delivery of services). The simulation model accounts for both hardware, as well as software failures, modeling their occurrences and detection, as well as recovery and masking (redundancy) mechanisms [1].

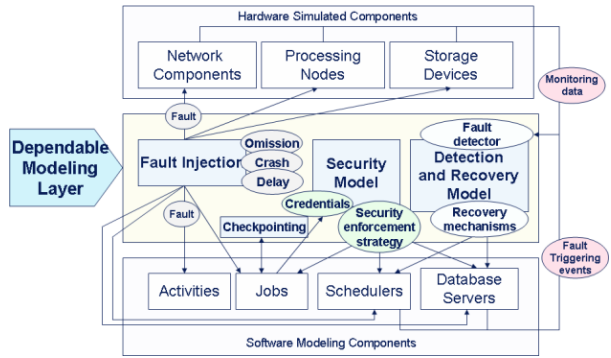

*Figure 3. The dependable simulation model and its components.*

## 3. Fault tolerance model

At hardware level different distributed components can be modeled as failing: the processing unit, the network connectivity as well as the storage devices. At software level we consider the faults occurring in a middleware component (the scheduler behavior could be erroneous, the database server could return wrong values, etc.) or in the higher-level distributed application (for example the jobs could fail to return correct results). For all the modeled components being considered by the simulation model we added specific mechanisms to inject faults. The injection of faults affects primarily the behavior of the components. In this way we are able to model different faults: crash faults, omission faults, time faults, as well as Byzantine faults [1]. In order to cover all possible faults, the model includes the use of any of two mechanisms, as follows.

In the first approach the user input into the simulation model values for the MTTF (mean time to failure) parameter in case of the various components involved in a specific simulation experiment. This parameter represents the basis for simulating the haphazardness stimuli coming from external and internal sources of influence that affect the characteristics of the modeled components and is seen as a probability measure of fault occurrence that is supposedly computed for each component prior to its deployment in the system. For modeling the fault injection mechanisms we use the MTTF together with a mathematical probability distribution (such as binomial, Poisson, Gaussian, standard uniform, etc). At random intervals of time, a component can therefore experience faulty behavior (failures), as well as possible recovery. Regarding the failures, a component can experience complete crash (the component will not recover anymore and will not be accessible by modeled entities), omissions (the network model will deliver only partial messages for example) or delays (the component will experience modeled delays).

The second proposed approach considers a completely random occurrence of fault events, without the user specifying any input value. This is useful in modeling the most disruptive faults, the Byzantine failures that occur arbitrary in the simulated system, such as in the case of transient hardware faults for example: they appear, do the damage, and then apparently just go away, leaving behind no obvious reason for their activation in the first place. For these types of errors the simulation model allows resuming the normal behavior of the affected component, as it is usually enough to allow successful completion of its execution.

The fault injection mechanisms are used together with various fault detection and recovery mechanisms. For that the model is augmented with a monitoring component. The monitoring component is responsible with receiving data of interest (such as fault occurrence triggered events), taking actions to update the state of the distributed system being modeled and possible inform interested components of the events occurrences. This component can track resource faults in the system and generate appropriate controlling actions. For example, based on various fault injection mechanisms, we can trigger the generation of a crash of a processing unit. The trigger translates in the generation of a special simulation event that is received by the monitoring component. Next the monitoring unit removes the processing unit from the modeled system (it will no longer be visible and all its on-going tasks will be forced to stop – the state update action) and inform all interested components of the occurrence of the event. In this approach a scheduler that is interesting in monitoring the state of the processing unit on which it deployed a task for execution would register with the monitoring component and, upon triggering of the crash event, will be informed of the failure so that to take the appropriate correction actions). We actually included in the model (and in the implementation in MONARC) an implementation of a DAG scheduling algorithm that is capable of taking appropriate rescheduling decisions when faults occur. Such a component is useful for example when evaluating different recovery schemes for distributed systems.

In this model an omission fault example is the simulation of a crushed network link, where the monitoring unit is responsible with generating corresponding interrupt events for the still-running tasks that were using that modeled link.

For modeling timing faults the monitoring unit also plays an important role. In the simulation model the boundaries of the action (start of the execution of a task, termination of a task, etc.) are modeled using simulation events. When the monitoring unit receives a timing fault triggering event it will simply modify termination events so that to be triggered at a later time in the future. In this way we simply change the default behavior so that a task will not end when it was supposed to end, but sometimes later in the future. A modeled fault-tolerant software component could then use timing checks (deadlines) to look for deviations from the acceptable behavior. For example, the scheduler also implements a fault-tolerant mechanism, according to which whenever a new job is submitted the scheduler also produces a special simulation event that triggers when the timeout occurs (where by timeout we mean an amount of time dependant on the user specification that will be triggered if the job fails to return results in due time). The scheduler will then be interrupted by one of two actions: either the job finishes or the timeout event occurs (which one happens faster in the simulation timeline). The same mechanisms are implemented by the network simulation model, a job being informed if a transfer failed to finished in a specified amount of time (possible due to network congestion for example) in order to consider taking appropriate measures (such as canceling the transfer for example).

We state that the described extension to MONARC's simulation model is useful for testing both reactive and proactive fault tolerance existing techniques [13]. In case of the reactive fault tolerant systems the fault monitoring systems is informed after the detection of the failure in order to start the corresponding recovery processes. The predictive (proactive) fault tolerant systems predict a fault before its occurrence. Predictive systems are based on the assumption that faults do show some disruptive effect on the performance of the system. In our approach the user can evaluate the performance of the distributed system in the presence of various predictive fault tolerance techniques by augmenting the monitoring component with various prediction algorithms.

In order to allow the evaluation of a wide-range of dependability technologies, the simulation model also includes the mechanisms to allow the modeling of check-pointing or logging of the system's state. These mechanisms are implemented based on the simulation events and the state of the objects used to simulate the modeled components. For that the job interface provides a method that, when called, results in the saving of the serialized objects as well as the state of the simulation on the local disk storage. This is useful in experiments dealing with both static and dynamic check-pointing strategies.

The simulation model also includes the evaluation of various replication and redundancy mechanisms. The replication provides mechanisms for using multiple identical instances of the same system or subsystems and choosing the result based on quorum. The simulation model allows the simulation of DAG distributed activities. This construction is also useful in modeling the replication of the jobs, where the same job would be executed on multiple processing units and another job is used to receive the outputs and select the correct result.

Redundancy results were demonstrated by the experiments described in [8]. In the experiments we describe how the simulation model already deals with replicating database storages. We also added replication mechanism to in case of the simulated jobs and scheduler as well.

## 4. Security model

Distributed systems are more vulnerable to security threats than traditional ones due to aspects such as the need for distributed access control, remote access to resources or the wide spread of resources, located in different administrative domains. We present the extension to the original MONARC's model designed to consider the evaluation of security in distributed systems.

In the case of distributed systems the main security threats try to exploit the weaknesses of the protocols and operating systems underneath them, but also the ones exploiting higher levels such as the ones implying attacks over databases, file sharing or multimedia applications, etc. Therefore, a complete security model must consider security aspects ranging from confidentiality of the data, authentication, non-repudiation, data integrity, access control, as well as key management [14].

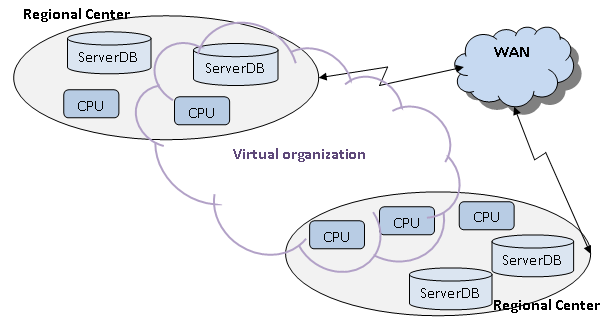

*Fig. 4. The modeling of virtual organizations.*

The starting point in designing the security model was the one available in the real-life middleware Globus Toolkit [15]. The model allows the definition of virtual organizations (VOs), each one can being able to share resources belonging to different regional centers. The model assumes authentication based on X.509 certificates, as in case of Globus. The users submitting jobs must therefore present a valid certificate used to verify their identity within the relation with the different entities throughout the modeled system.

The security model also allows mutual authentication, much like in case of GSI [15], for the initial phase of the communication. The model considers the default SSL mutual authentication protocol. But the security model is extensible, allowing for example also the use of unidirectional authentication, or the protection of messages using a user-defined SSL-like encryption protocol. The access authorization to the resources of the system is also part of the model, being simulated through the use of security policies at the level of components belonging to VOs.

The main mechanisms considered for the security model are: the possibility to create virtual organizations, security policies for VOs, the use of authentication, authorization, the possibility to protect messages using cryptographic protocols, to secure data transfer or to filter traffic. The proposed model therefore describes a wide range of security solutions, such as GSI, PKI, SSL, cryptographic solutions, etc., and is adequate for modeling attach ranging from DoS to detecting attacks using cryptographic messages or authentication and authorization protocols, as well as the modeling of possible reactions to such attacks.

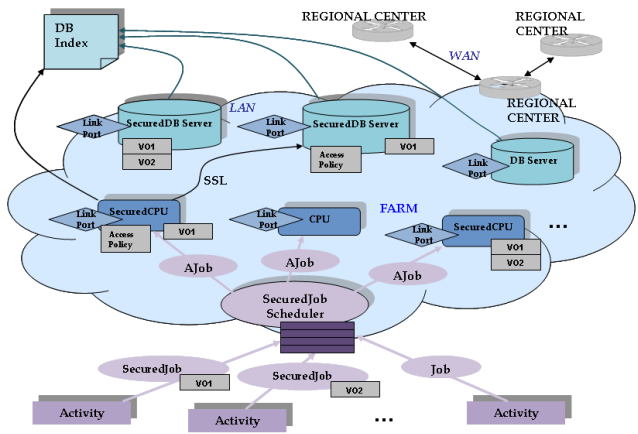

*Fig. 5. The components of the security model.*

In order to implement the model within MONARC all basic components were extended (Figure 5): the processing CPU unit, the job, the database server, the job scheduler – they were all added the mechanisms for authentication, access control and job scheduling according to the restrictions imposed by the VO where the jobs should be executed.

The secured job identify the job submitted for execution within a VO (the default mechanism is based on the use of certificate, but other user-defined mechanisms can also be used). For that we added the possibility to define the VO, as well as added the mechanisms for authentication within the system (and delegation as well).

The secured job scheduler schedules jobs for execution based on access control policies defined using various mechanisms (such as the mapping of certificates versus VOs).

The processing unit was also extended such that to specify the VO to which it belongs. By default for each VO the secured processing unit has an associated security policy and access control method. We actually evaluated in different scenarios access control methods based on the required processing power or memory needed for successful processing. When a job is submitted for execution on a processing unit a series of tests for the determination of identity and authorization, as well as the existence of proper access rights are executed. A regional center can have multiple processing units, shared within one or more virtual organization, as well as processing units that do not belong to any virtual organization (not shared within the cloud).

The database server was also extended to support the sending and receiving of secured messages, as well as various access control policies defined for the access to the data locally stored. In this case the database server maintains a mapping between virtual organizations and access policies. When one tries to access the data on the server a series of tests are being automatically executed, having the role of determining the identity and access roles and, based on the type of operation and data, determine if the user has the right to access the secured data or not. Such a mechanism can be used to model various types of security access solutions, from distributed file systems to user-oriented operating system access control policies.

Also the security model considers the use of cryptographic protocols (such as SSL) to encrypt communication. Such mechanisms are especially useful for ensuring security and confidentiality of the data transferred being model within a simulation scenario. The user has also the possibility to define its own data transfer security protocol, as well as cryptographic algorithms being used.

Finally, in the model we added the possibility of traffic filtering (like in the case of firewalls or routers) within each component. The filtering is based on rules defined statically or dynamically (at simulation time) by the user. To complete the model, we also defined a series of patterns for attacks (such as DoS), useful for the evaluation of various user-defined security mechanisms.

## 5. Results

In a first series of experiments we evaluated the capability of the simulation model to cope with fault tolerance solution in different parts of a scenario. We first evaluated the capability of producing and recovering from failures in the network layer. For these experiments we considered a scenario composed of four regional centers all connected to each others. Within each regional center we considered the existence of a LAN, while between the WANs connecting the regional centers we assumed the existence of two modeled routers.
The network traffic was generated so that the packets would traverse at least one router. In the experiments faults were generated in the LANs, as well as in the routers. We were particularly interested in the correlation between the number of lost packets in the network and the probability coefficients of failures occurrences in the intermediary components.

The retransmission is accomplished by the upper-level protocols – TCP in this case. In the first experiment we simulated the failure of the link port interfaces. When the interface experiences transient faults, after several attempts, the messages are correctly delivered. The results in case of permanent crashes are presented in Figure 6.

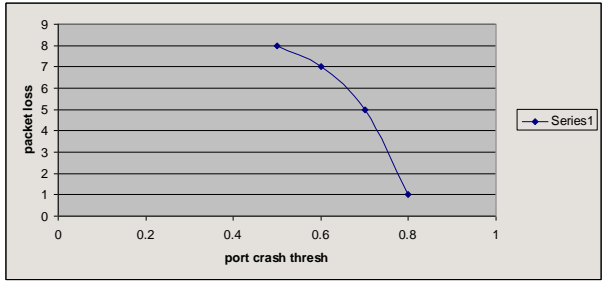

*Fig. 6. The influence of modeled failures over a network transmission in a LAN.*

We next experimented with faults occurring in the routers. The failure of one routers leads to an interruption in the communication. As in the previous experiment, the transient failures only leads to an increase in the time needed to transfer the messages, but eventually the transmission ends correctly. The influence of crash failures is presented in Figure 7.

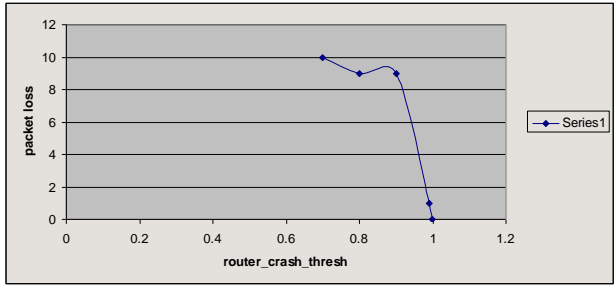

*Fig. 7. The influence of modeled failures in a router over a network transmission.*

A first series of simulation experiments evaluated the fault-tolerant scheduling algorithm for DAGs that is part of the model. The experiments considered the case of several complex DAG dependent tasks that were submitted for execution, and the cases when faults occur or not. We first analyzed the report between the finalized versus submitted tasks for both cases.

We conducted these experiments on the conditions and algorithms proposed in [12]. The results are presented in Figure 8.

In figure 8 the difference between the submitted jobs and the finalized ones represents the number of jobs that were successfully rescheduled (when faults occurred).

An experiment designed to evaluate the security model considered the case of two regional centers (Figure 9) sharing several processing units and a database server within a virtual organization. The purpose of this scenario was to demonstrate the functionality of an access policy within the secured database server and to identify flows within the model.

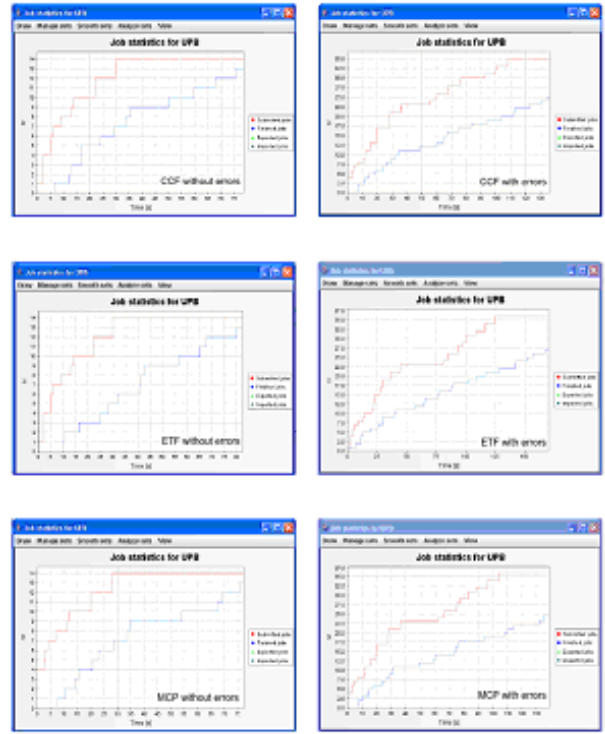

*Fig. 8. Results obtained when evaluating the fault-tolerant scheduling algorithms – the case of running the CCF, ETF and MCP scheduling algorithms [12].*

For this simulation experiment we defined two jobs: one requests the creation of a database and writes data in it and the other one connects to the server and requests the data matching a specific pattern.

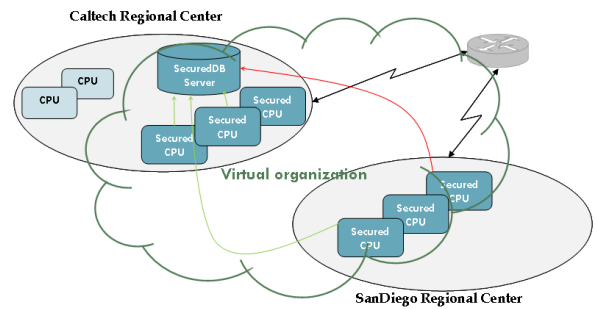

*Fig. 9. The scenario configuration.*

We associated a security policy resembling the UNIX file access policies to the database server belonging to the VO. We next considered that the members of the VO have read and write rights over the database server. Any get operation will be ignored and the operation is considered an implicit attack on the database server.

The experiment consisted in the insertion of many jobs of the types previously presented. The obtained results, presented in Figure 10, demonstrate that, as expected, during an attack the throughput increases, in contrast with the initial conditions of the experiments. Also, the number of received connections increases during an attack. The results demonstrate the validity of the proposed security model, as they are well mapped with the analytical results expected from the experiment. We also conducted a number of other experiments, trying to evaluate the components proposed within the security model, ranging from securing communication to imposing access control at virtual organization level.

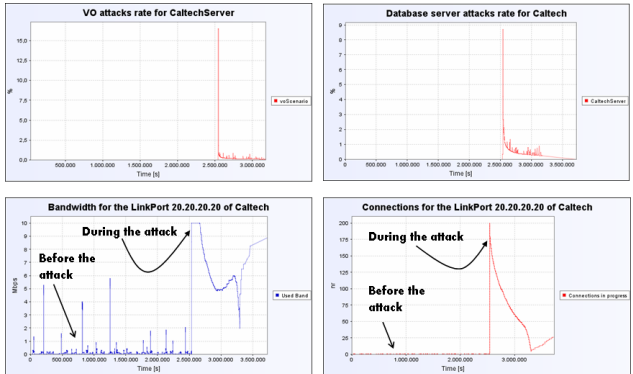

*Fig. 10. Results obtained in case of the security simulation experiment.*

## 6. Conclusions

As society increasingly becomes dependent of distributed systems (Grid, P2P, network-based), it is becoming more and more imperative to engineer solutions to achieve reasonable levels of dependability for such systems. Simulation plays an important part in building and evaluating dependable distributed systems.

In this paper we presented a new simulation model designed to evaluate the dependability in distributed systems. This model extends the MONARC simulation model with new capabilities for capturing reliability, safety, availability, security, and maintainability requirements. The model has been implemented as an extension of the multithreaded, process oriented simulator MONARC, which allows the realistic simulation of a wide-range of distributed system technologies, with respect to their specific components and characteristics.

The extended simulation model includes the necessary components to inject various failure events, and provides the mechanisms to evaluate different strategies for replication, redundancy procedures, and security enforcement mechanisms, as well. The results obtained in simulation experiments presented in this paper probe that the use of discrete-event simulators, such as MONARC, in the design and development of distributed systems is appealing due to their efficiency and scalability.